\documentclass[preprint,12pt]{elsarticle}

\usepackage{amssymb}

\journal{Physics Letter B}

\begin{document}

\begin{frontmatter}



\title{Electric conductivity of hot and dense nuclear matter}


\author[inst1]{Joseph Atchison}

\affiliation[inst1]{organization={Abilene Christian University},
            addressline={1600 Campus Ct}, 
            city={Abilene},
            postcode={79601}, 
            state={Texas},
            country={USA}}

\author[inst2,inst3]{Yiding Han}
\author[inst2]{Frank Geurts}
\affiliation[inst2]{organization={Rice University},
            addressline={6100 Main St}, 
            city={Houston},
            postcode={77005}, 
            state={Texas},
            country={USA}}
            
\affiliation[inst3]{organization={Baylor College of Medicine},
            addressline={1 Baylor Plz}, 
            city={Houston},
            postcode={77030}, 
            state={Texas},
            country={USA}}

\begin{abstract}
Transport coefficients play an important role in characterising hot and dense nuclear matter, such as that created in ultra-relativistic heavy-ion collisions (URHIC). In the present work we calculate the electric conductivity of hot and dense hadronic matter by extracting it from the  electromagnetic spectral function, through its zero energy limit at vanishing 3-momentum. We utilise the vector dominance model (VDM), in which the photon couples to hadronic currents predominantly through the $\rho$ meson. Therefore, we use hadronic many-body theory to calculate 
the $\rho$-meson's self-energy in hot and dense hadronic matter, by dressing its pion cloud with $\pi$-$\rho$, $\pi$-$\sigma$, $\pi$-$K$, N-hole, and $\Delta$-hole loops. We then introduce vertex corrections to maintain gauge invariance. Finally, we analyze the low-energy transport peak as a function of temperature and baryon chemical potential, and extract the conductivity along a proposed phase transition line.
\end{abstract}

\begin{graphicalabstract}
\vspace*{\fill}

\noindent
\makebox[\textwidth]{\includegraphics[width=50pc]{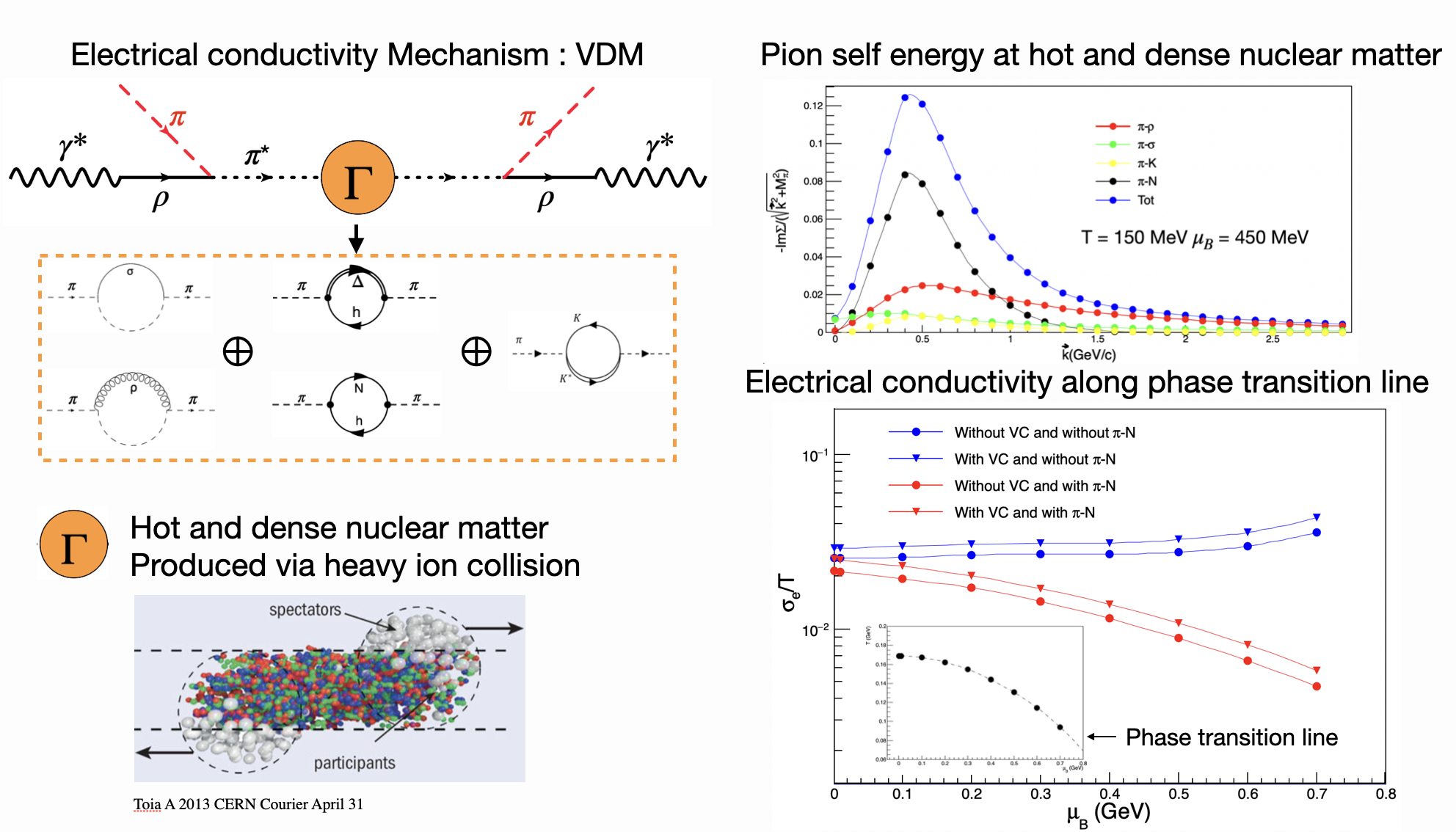}}%

\vspace*{2cm}
\end{graphicalabstract}

\begin{highlights}
\item We calculate the EM-spectral function at finite temperature and density, along the QCD phase transition line.
\item EM-currents couple to the medium through the vector mesons, with the dominate contribution coming from the $\rho$ meson.
\item The $\rho$ meson's medium interactions are quantified through the $\rho$'s self-energy $(\Sigma_{\rho})$.
\item The EM-spectral function's transport peak is generated by the absorption of a virtual $\rho$ by a thermal pion.
\item Thermal pions act as charge carriers. The pion's medium interactions are generated by scatterings with other thermal pions, kaons, and nucleons.
\item The pion interactions are introduced through the pion's self-energy $(\Sigma_{\pi})$, and interactions with nucleons are found to greatly increase the pion's width at finite baryon chemical potential $(\mu_{B})$.
\item At $\mu_{B}=0$ the width decreases and the conductivity increases at temperature falls. At finite $\mu_{B}$ nucleon interactions greatly influence the transport peak as its width increases and the conductivity decreases as $\mu_{B}$ rises.
\end{highlights}


\end{frontmatter}


\section{Introduction}\label{sec1}
Ultra-relativistic heavy-ion collisions (URHIC) are a mainstream method for simulating the hot and dense conditions of the early universe.These collisions produce Quark-Gluon Plasma (QGP), a deconfined phase with quarks and gluons that likely existed microseconds after the Big Bang~\cite{bigbang}. This phase may also explain the deconfined behaviors observed within the inner core of massive stellar objects~\cite{stellar}. As the system cools, the QGP undergoes a transition into a hadronic gas across the QCD phase transition line. Experiments at the Super Proton Synchrotron (SPS), the Relativistic Heavy-Ion Collider (RHIC), and the Large Hadron Collider (LHC) operate over a range of center-of-mass energies, probing the QCD phase diagram at various temperatures and baryon chemical potentials ($\mu_{B}$). In this paper we examine the properties of the phase transition line from the hadronic side.

Two approaches to studying hot and dense nuclear matter include the observation and description of electromagnetic probes, and the characterization of the medium with transport coefficients. Dileptons and photons are produced throughout the evolution of the nuclear fireball with little final state interaction, providing a probe of the electromagnetic spectral function ($\rho_{EM}$) in the medium~\cite{LandoltBornstein}. The production of low-mass dilepton and photon radiation has been well described at multiple experiments~\cite{Rapp:1999us,Salabura:2020tou,Tripolt:2022hhw}. Transport coefficients detail the long-wavelength properties of the medium, describing the flow of conserved charges throughout the collision. In this paper, we calculate the electric conductivity ($\sigma_{el}$), which can be extracted from the zero-energy limit of $\rho_{EM}$ at vanishing 3-momentum~\cite{Feinberg:1976,McLerran:1985,Moore:2006}. A close relationship exist between electromagnetic probes and the conductivity, as both are proportional to $\rho_{EM}$. In fact, future plans to measure very-low mass and momentum dileptons at the Schwer-Ionen Synchrotron (SIS), RHIC, and the LHC have potential to probe the conductivity's transport peak~\cite{HADES,STAR:iTPC,ALICE:2022wwr}. For example, recent work combining hadronic many-body theory with nonperturbative QGP found that a broadening transport peak will induce an enhancement in low-mass dilepton spectra that may be observable in future experiments at the LHC~\cite{Rapp:2024}. These initiatives encourage the study of the conductivity, especially considering the significant variation in calculations using different methodologies~\cite{Greif:2016,Huot:2006,Fraile:2006,Finazzo:2014,Aarts:2015,Brandt:2013,Brandt:2016,Amato:2013,Kadam:2019,Ghosh:2017,Ghosh:2018,Hammelmann:2018ath,Atchison:2023}. 

In this article we build on the work of~\cite{Atchison:2023}, which calculated the conductivity of hot pion matter in a hadronic quantum many-body framework. 
It was found that pions play an important role in the formation of the transport peak, because higher mass particles are suppressed at low energies. For this reason, pions are expected to acts as the primary charge carrier in hot and dense hadronic matter. However, pion scattering off larger medium particles, most notably nucleons, should significantly increase the pion's width and decrease the conductivity~\cite{Urban:1998,Urban:1999im}. Therefore, we extend~\cite{Atchison:2023}'s calculation to hot and dense hadronic matter, by incorporating previously evaluated hadronic interactions, such as $\pi N$ scattering from Refs. \cite{Urban:1998,Urban:1999im}, and $P$-wave pion scattering with thermal kaons~\cite{RappDiL,Rapp:1999prc}. Pion scattering is incorporated through the pion self-energy ($\Sigma_{\pi}$), which quantifies the pion's medium interactions. In order to maintain gauge invariance, it is necessary for $\rho_{EM}$ to be 4-dimensionally transverse. Transversality can be assured by the construction of vertex corrections that satisfy the Ward Identities. While it is straight forward to implement the vertex corrections from~\cite{Urban:1999im}, the corrections derived in~\cite{Atchison:2023} are more numerically cumbersome. Therefore, for $\pi\pi$ and $\pi K$ scattering, we introduce effective vertex corrections, which exactly maintain the Ward Identities. Here, we calculate $\rho_{EM}$ and $\sigma_{el}$ along the phase transition line in hot and dense hadronic matter. Results are given with and without vertex corrections, and the significance of each scattering processes is examined at various temperatures and baryon chemical potentials.

This paper is organized into six sections: in Section \ref{sec2} we review the relation of $\rho_{EM}$ to the $\rho$ meson propagator, and introduce the $\rho$ meson's self-energy ($\Sigma_{\rho}$). Section \ref{sec3} addresses the pion's medium interactions, including scatterings with other pions, kaons, and nucleons. In Section \ref{sec4} we introduce the Ward Identities and detail efforts to maintain gauge invariance by constructing vertex corrections. Section \ref{sec5} provides our numerical results for the EM spectral function. We emphasize the low-energy transport peak, and extract the conductivity along the phase transition line. Finally, we summarize and discuss future work in \ref{sec6}.

\section{EM spectral function in hadronic matter}\label{sec2}

One can relate the thermal dilepton emission rate in hot and dense nuclear matter to $\rho_{EM}$ through~\cite{Feinberg:1976,McLerran:1985}:
\begin{eqnarray}\label{dilepton_rate}
\frac{dR_{l+l-}}{d^4q}=\frac{\alpha_{\textrm{EM}}^2}{2\pi^3 M^2}f(q_0,T)\rho_{\textrm{EM}}(M,q,T,\mu_B) \ ,  
\end{eqnarray} 
where $f$ denotes the Bose-Einstein distribution, $M^{2}$ is the dilepton's invariant mass, and 
$\alpha_{\textrm{EM}}=\frac{e^{2}}{4\pi}$ the fine-structure constant. The spectral function is related to the EM current-current correlation function ($\Pi_{EM}$) through  $\rho_{\rm EM}= -2 \ {\rm Im}\Pi_{\rm EM}$, where we have taken the polarization average $\Pi_{\rm EM}\equiv g_{\mu\nu} \Pi_{\rm EM}^{\mu\nu}/3$. The conductivity is in turn related to the spacial components of the EM spectral function via:
\begin{eqnarray}
\label{conductivity}
\sigma_{el}(T)=(-e^{2}/3) \lim_{q_0 \to 0} [\Pi_{\textrm{EM}}^{ii}(q_0,\vec{q}=0,T)/q_{0}] \ .
\end{eqnarray} 
From Eq. \ref{conductivity} one sees that $\rho_{EM}$ must approach zero linearly in order to obtain a finite/non-zero conductivity.

The EM correlator is derived from the EM current ($j_{\textrm{em}}^{\mu}$) through~\cite{Sakurai}:
\begin{eqnarray}\label{correlator_to_current}
\Pi_{\textrm{\textrm{EM}}}^{\mu\nu}(q_{0},\vec{q})=-i\int d^{4}x e^{i q\cdot x}\Theta (x_{0})\langle[j_{\textrm{em}}^{\mu}(x),j_{\textrm{em}}^{\nu}(0)]\rangle.
\end{eqnarray}
For large invariant mass $j_{\textrm{em}}^{\mu}$ is well described by a quark continuum, while for invariant mass less than approximately 1 GeV, $\Pi_{EM}$ couples to the medium primarily through the neutral vector mesons. The complementary nature of the two regimes is seen if one expresses the vector mesons in terms of there quark content:
\begin{eqnarray}
\label{j_quark}
j_{\textrm{em}}^{\mu}=\frac{1}{\sqrt{2}}\bar{\psi}\gamma^{\mu}\psi \Bigg[\frac{\bar{u}u-\bar{d}d}{\sqrt{2}}+\frac{1}{3}\frac{\bar{u}u+\bar{d}d}{\sqrt{2}}-\frac{\sqrt{2}}{3}\bar{s}s\Bigg],\\
\label{j_vector}
j_{\textrm{em}}^{\mu}(M\leq 1\, \textrm{GeV})=\frac{m_{\rho}^{2}}{g_{\rho}}\rho^{\mu}+\frac{m_{\omega}^{2}}{g_{\omega}}\omega^{\mu}+\frac{m_{\phi}^{2}}{g_{\phi}}\phi^{\mu}.
\end{eqnarray}
From Eq. \ref{j_quark} it is apparent that the $\omega$ and $\phi$ mesons are suppressed relative to the $\rho$. In fact, for small invariant mass one can approximate $\textrm{Im}\Pi_{EM}$ in terms of only the $\rho$ meson's spectral function ($\textrm{Im}D_{\rho}^{\mu\nu}$)~\cite{LandoltBornstein,Sakurai}:
\begin{eqnarray}
\label{VDM}
\textrm{Im}\Pi_{\textrm{\textrm{EM}}}^{\mu\nu}&\approx& \frac{(m_{\rho}^{(0)})^{4}}{g_{\rho}^{2}}\textrm{Im}D_{\rho}^{\mu\nu} \ ,
\end{eqnarray} 
The $\rho$'s interactions are quantified in the $\rho$ meson self-energy ($\Sigma_{\rho}^{\mu\nu}$), which can be resummed into $D_{\rho}$. In the proceeding sections, we review the components the vacuum $\rho$ self-energy, and detail the contributions to $\Sigma_{\rho}^{\mu\nu}$ in medium.

We adopt the rho self-energy described in ~\cite{Atchison:2023,Urban:1998}, which began with the vacuum $\Sigma_{\rho}^{\mu\nu}$ described in ~\cite{HerrmannGI}. The self-energy was then calculated at finite temperature within the imaginary time formalism using techniques from Refs.~\cite{Fetter,Rapp:1996ym} to obtain: 
\begin{eqnarray}
\label{pipi}
\Sigma_{\rho}'^{\mu\nu}(q)&=&\frac{1}{2} \int \frac{d^3k}{(2\pi)^3} \int_{-\infty}^{\infty}\frac{dvdv'}{\pi^{2}}\frac{\Gamma_{\mu\,3ab}^{(3)}(-k,q)\Gamma_{\mu\,3ba}^{(3)}(q-k,-q)}{q_{0}-v-v'+i\epsilon}\nonumber\\
&&\times\textrm{Im}[D^{0}_{\pi}(v,\vec{k})]\textrm{Im}[D^{0}_{\pi}(v',\vec{k}+\vec{q})](1+f(v)+f(v'))+\textrm{I}_{\textrm{Tad}}  \ ,
\end{eqnarray}
where
\begin{eqnarray}\label{tad}
\textrm{I}_{\textrm{Tad}}&=&\frac{-i}{(2\pi)^{4}} \int d^{3}k\int_{-\infty}^{\infty}dv\textrm{Im}[D^{0}_{\pi}(v,\vec{k})]\Gamma_{\mu\nu\,aa33}^{(4)}(-k,q) (1+f(v)) \ ,
\end{eqnarray}
$D_{\pi}^{0}$ is the free pion propagator, $f(\nu)=1/(e^{(\nu-\mu_{B})/T}-1)$, $\Gamma_{\mu\,3ab}^{(3)}$ is the vacuum $\rho\pi\pi$ vertex, $\Gamma_{\mu\nu\,aa33}^{(4)}$ is the vacuum $\rho\rho\pi\pi$ vertex, and $k^{0}=\nu$. The pion propagator and the vacuum vertices are given by: $D^{0}_{\pi}(k)=1/(k^{2}-m_{\pi}^{2}+i\epsilon), \Gamma_{\mu\,abc}^{(3)}=g_{\rho}\epsilon_{cab}(2k+q)_{\mu}, and \Gamma_{\mu\nu\,abcd}^{(4)}=ig_{\rho}^{2}(2\delta_{ab}\delta_{cd}-\delta_{ac}\delta_{bd}-\delta_{ad}\delta_{bc})g_{\mu\nu}$.

Regularization with a standard form-factor would violate current conservation, thus we implement a Pauli-Villars scheme, as detailed in~\cite{Urban:1998,HerrmannGI}. We adopt the values of $g_{\rho}=5.9$, $m_{\rho}^{(0)}=853 \, \textrm{MeV}$, and $\Lambda_{0}=1 \, \textrm{GeV}$, which were obtained by fits 
to the $P$-wave $\pi\pi$ phase shifts and pion electromagnetic form factor~\cite{Urban:1998}.

The vacuum is composed of the two interactions. The $\pi\pi$-loop is responsible for generating the $\rho$'s vacuum width, through two pion decay, while the tadpole loop is purely real and is required for $\Sigma_{\rho}^{\mu\nu}$ to be 4-dimensionally transverse. These two interactions account for the $\rho$'s so called "pion cloud" in vacuum. At finite temperature, $\textrm{Im}\Sigma_{\rho}^{\mu\nu}$ can be further decomposed into the Unitary and Landau cuts. As demonstrated in Ref.~\cite{Atchison:2023}, the Unitary cut describes $\rho\rightarrow\pi\pi$ decay, as well as its thermal Bose enhancement; the Landau cut describes the absorption of a virtual $\rho$ by a thermal pion, and generates the transport peak. 

In medium Lorentz invariance is broken, resulting in the $\rho$ propagator splitting into transverse and longitudinal modes for finite 3-momentum~\cite{Gale}. However, for $\vec{q}=0$ the transverse and longitudinal modes become degenerate, allowing one to write the conductivity in terms of only the transverse mode:
\begin{eqnarray}
\label{Trans_conductivity}
\sigma_{el}=\frac{e^2}{g_{\rho}^{2}} \lim_{q_0 \to 0} \textrm{Im}\big[\frac{-(m^{0}_{\rho})^{4}}{q_{0}^2-(m_{\rho}^{0})^{2}-\Sigma_{\rho}^{T}(q_{0},\vec{q}=0)}\big] \ ,
\end{eqnarray}
where $\Sigma_{\rho}^{T}$ is the transverse projection of $\Sigma_{\rho}^{T\mu\nu}$.

Therefore, in order to determine $\sigma_{el}$, one must calculate $\Sigma_{\rho}^{T}$, in the limit of zero energy and momentum. At finite temperature and baryon chemical potential the $\rho$ can interact with the medium directly, through resonant interactions with nucleons and mesons, or indirectly, through its pion cloud. The direct interactions were calculated in Refs.~\cite{RappDiL,Rapp:1999prc,ANP25}, where current conservation required the pertinent self-energies be proportional to the square of the $\rho$'s energy, 3-momentum, or invariant mass. One can see from Eq. \ref{Trans_conductivity} that these interactions cannot contribute to the conductivity. Furthermore, Ref. \cite{Urban:1998} demonstrated that to first order in the nonrelativistic expansion nucleons do not contributed to the EM-current. Therefore, we only consider direct $\rho$ interactions with pions, which is equivalent to assuming the pion acts as the only charge carrier. In the next section, we quantify the pion's medium interactions in terms of pion self-energies ($\Sigma_{\pi}$), which are incorporated into $\Sigma_{\rho}^{T}$ through resummation into the pion propagator.

\section{Pion interactions in medium}\label{sec3}
In this section we calculate the pion's interactions at finite temperature and baryon chemical potential. These interactions are expressed through $\Sigma_{\pi}$ and resummed into $D_{\pi}$: $D_{\pi}(k_{0},\vec{k})=1/(k^{2}-m_{\pi}^{2}+\Sigma_{\pi}(k,T,\mu_{B}))$. We consider resonant scattering with other pions, nucleons, and kaons.

For $\pi\pi$ scattering we use the pion self-energies used in Ref.~\cite{Atchison:2023}, where the S and P-wave interactions were calculated, assuming s-channel dominance, through $\sigma$ and $\rho$ resonant scattering, respectively.

For $\pi N$ scattering we adopt the self-energies calculated in Refs.~\cite{Urban:1998,Urban:1999im}. These self-energies quantify the absorption of a in-medium nucleon by a pion and the subsequent formation of a nucleon or delta resonance, i.e. a nucleon-hole (Nh) or delta-hole ($\Delta$h) excitation.

\begin{figure}
\centering
\includegraphics[width=19pc]{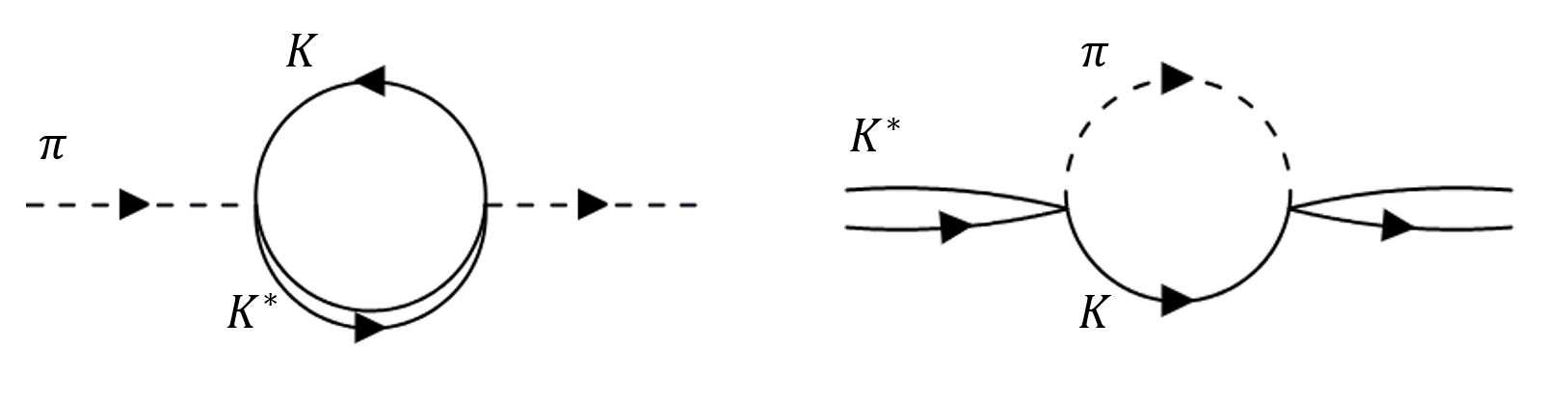}
\includegraphics[width=20pc]{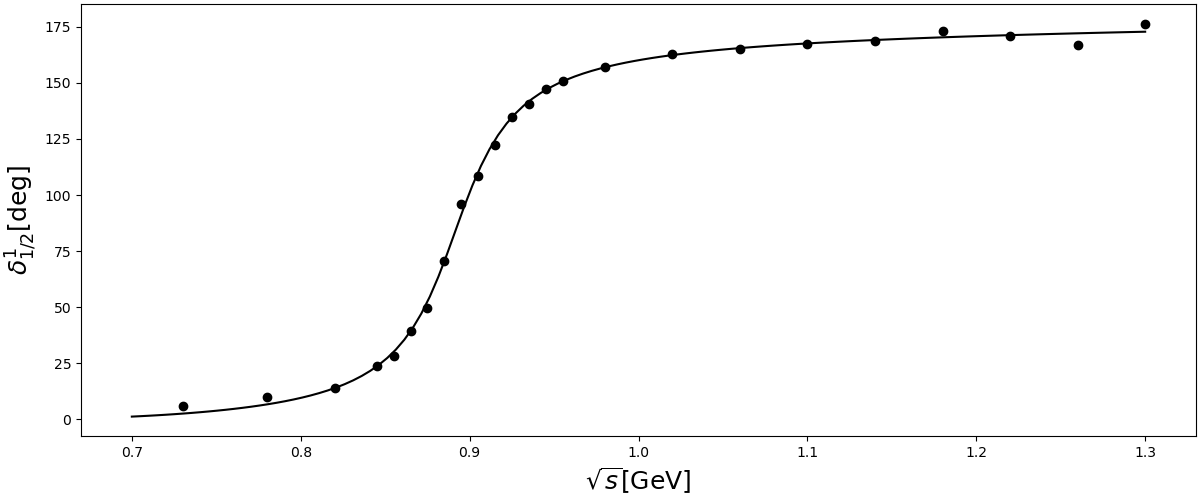}
\caption{\label{kaon_self}Top: The pion self-energy resulting from resonant scattering of a cloud pion with a thermal Kaon through an intermediate $K^{*}$ (left). $K^{*}$ vacuum self-energy resulting from $K^{*}\rightarrow \pi K$ decay (right). Bottom: Fit to the isospin-1/2 P-wave $\pi K$ scattering phase shift via a $K^{*}$ resonance, compared to experimental data~\cite{Estabrooks:1977xe}}
\end{figure}

For $\pi K$ scattering we consider P-wave scattering through a $K^{*}(892)$ resonance (upper left panel of figure \ref{kaon_self}). We start from the effective $\pi K$ interaction Lagrangian presented in~\cite{Li:2003}:
\begin{eqnarray}\label{Kaon_Lagrangian}
\mathcal{L}_{\pi K K*}&=&ig_{K}\big\{\big[(\partial_{\mu}\bar{K})\vec{\tau}K^{*\mu}-\bar{K}^{*\mu}\vec{\tau}(\partial_{\mu}K)\big]\cdot\vec{\pi}\nonumber\\
&&-\big[\bar{K}\vec{\tau}K^{*\mu}-\bar{K}^{*\mu}\vec{\tau}K\big]\cdot(\partial_{\mu}\vec{\pi})\big\},
\end{eqnarray}
where $g_{K}$ is the $\pi K K^{*}$ coupling. The $\pi K K^{*}$ vertex can be derived from Eq. \ref{Kaon_Lagrangian}:
\begin{eqnarray}\label{Kaon_Vertex}
\Gamma_{K \, a}^{\mu} = g_{K}(2k+q)_{\mu}\tau_{a},
\end{eqnarray}
where $\tau_{a}$ are the Pauli spin matrices. We derive our $\pi K$ self-energy using Feynman rules:
\begin{eqnarray}\label{Kaon_selfenergy}
\Sigma_{\pi}^{K}&=&4 g_{K}^2\int \frac{d^3p}{(2\pi)^3}\int_{-\infty}^{\infty}\frac{dwdw'}{\pi^{2}}\Big[\textrm{Im}[D_{K}^{0}(w,\vec{p})]
\textrm{Im}[D_{K^{*}}(w',\vec{k}+\vec{p})]\nonumber\\
&&\times\Big(-(k-p)^{2}+\frac{(k^{2}-p^{2})^{2}}{w'^{2}-(\vec{p}+\vec{k})^{2}}\Big)\frac{\textrm{FF}^{2}_{K}(q_{\textrm{cm}}^2)(f(w)-f(w'))}{k_{0}+w-w'+i\epsilon}\Big]_{p_{0}=w},
\end{eqnarray}
where we have added a factor of 2 to account for scattering through $K^{+}$ and $K^{-}$. Additionally, we define the undressed vacuum K propagator $D^{0}_{K}(k)=1/(k^{2}-m_{K}^{2}-i\epsilon)$, and the vacuum $K^{*}$ propagator $D_{K^{*}}(q)=1/(q^{2}-(m^{(0)}_{K^{*}})^{2}-\Sigma_{K^{*}}(q))$ \ . Here $m^{(0)}_{K^{*}}$ is the $K^{*}$'s bare mass, and $\Sigma_{K^{*}}$ is the $K^{*}$'s vacuum self-energy, which we calculate assuming a 100\% branching ratio for $K^{*}\rightarrow \pi K$ decay (right panel of figure \ref{kaon_self}):
\begin{eqnarray}
\label{Kpi_loop}
\Sigma_{K^{*}}(q)&=&2 g_{K}^2 \int \frac{d^3k}{(2\pi)^3} \int_{-\infty}^{\infty}\frac{dvdv'}{\pi^{2}}\frac{(2k-q)^{\mu} (2k-q)^{\nu}}{q_{0}-v-v'+i\epsilon}\nonumber\\
&&\times\textrm{Im}[D^{0}_{\pi}(v,\vec{k})]\textrm{Im}[D^{0}_{K}(v',\vec{k}+\vec{q})]\textrm{FF}^{2}_{K}(k),\\
\label{Ks_vac_scalar_sig}
\Sigma_{K^{*}}^{\mu\nu}(k, T=0)&=&(-g^{\mu\nu}+\frac{k^{\mu}k^{\nu}}{k^{2}})\Sigma_{K^{*}}(k, T=0) \ .
\end{eqnarray}
Finally, the Kaon form factor is given by:
\begin{eqnarray}
\label{KaonFF}
\textrm{FF}_{K}(q^{2}_{\textrm{cm}})&=&\frac{\Lambda_{K}^{2}}{\Lambda_{K}^{2}+q_{\textrm{cm}}^2(p,k)},
\end{eqnarray}
where $q_{\textrm{cm}}$ is the pion momentum in the CM system of the collision. We fit our model to isospin-1/2 P-wave $\pi K$ scattering phase shift from Ref.~\cite{Estabrooks:1977xe}. The relevant phase shift is related to the $K^{*}$ propagator by:
\begin{eqnarray}
\label{PiK_PhaseShift}
\textrm{tan}\delta_{1/2}^{1}(E)&=&\frac{\textrm{Im}[D_{K^{*}}(E)]}{\textrm{Re}[D_{K^{*}}(E)]}.
\end{eqnarray}
Our fit is displayed in figure \ref{kaon_self}, and provides the values $g_{K}=4.97$, $m_{K^{*}}^{(0)}=976$ MeV, $m_{K}=494$ MeV, and $\Lambda_{K}=630$ MeV.

\begin{figure}

\includegraphics[width=16pc]{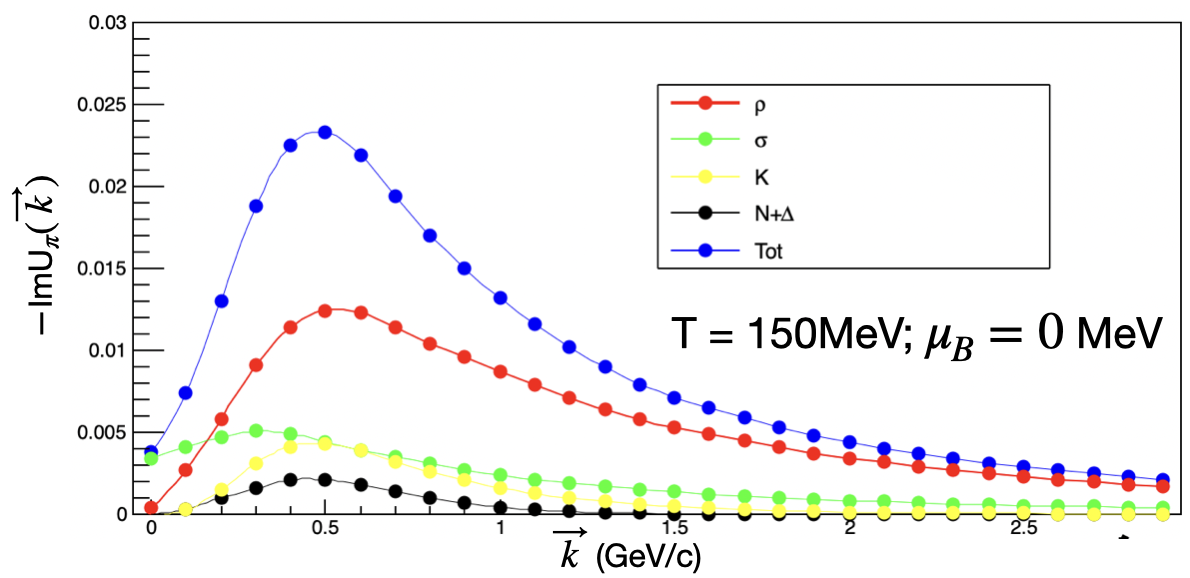}
\includegraphics[width=16pc]{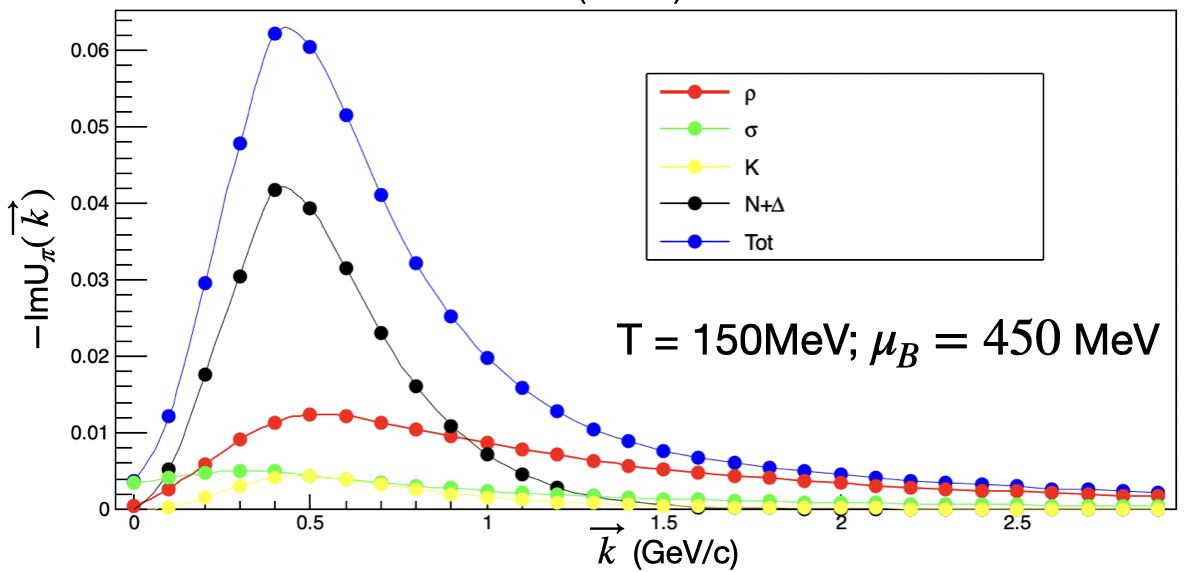}
\caption{The pion optical potential in hadronic matter at $T=150$ MeV. The contributions from scattering through various resonances are shown including: $\pi\pi$ scattering through a $\rho$/$\sigma$ (red/green) resonance, $\pi K$ scattering through a $K^{*}$ resonance (yellow), and $\pi N$ scattering through $N$ and $\Delta$ resonances (black). The total self-energy ($\rho$+$\sigma$+$K^{*}$+$N$+$\Delta$) is shown in blue. The left panel plots results for $\mu_{B}=0$, while the right panel plots results for $\mu_{B}=450$ MeV.}
\label{PionSelfTotal}
\end{figure}

The total in medium pion self-energy is given by the sum of the individual components, $\Sigma_{\pi}=\Sigma_{\pi}^{\rho}+\Sigma_{\pi}^{\sigma}+\Sigma_{\pi}^{N}+\Sigma_{\pi}^{K}$. It is convenient to plot the pion self-energy in terms of the optical potential $U_{\pi}(\vec{k})=\Sigma_{\pi}(\omega_{\pi},\vec{k})/(2\omega_{\pi}(\vec{k}))$: $U_{\pi}(\vec{k})=\Sigma_{\pi}(\omega_{\pi},\vec{k})/(2\omega_{\pi}(\vec{k}))$, where $\omega_{\pi}=\sqrt{\vec{k}+m_{\pi}}$. The optical potentials for the total pion self-energy and each component are plotted in figure \ref{PionSelfTotal} for $\textrm{T}=150$ MeV at zero and finite $\mu_{B}$.

For $\mu_{B}=0$, $\pi\pi$ scattering provides the dominate contribution, with S-wave scattering through the $\sigma$ meson being particularly important at low energy. The $\pi K$ interaction provides a modest increase, while the $\pi N$ interaction is heavily suppressed. At $\mu_{B}=450$ MeV  we include pion and kaon chemical potentials of $\mu_{\pi}=14.8$ and $\mu_{K}=2\mu_{\pi}$, which follow from Ref. ~\cite{Bebie:1991ij} (assuming a constant entropy density per pion of 3.7). The $\pi N$ interaction drastically increases at finite $\mu_{B}$ and dominates $\Sigma_{\pi}$. Alternatively, the inclusion of $\mu_{\pi}$ and $\mu_{K}$ result in only a small increase in $\Sigma_{\pi}$, as $\mu_{\pi}/\mu_{K}$ is small relative to the temperature. This demonstrates the importance of $\pi N$ scattering at large $\mu_{B}$'s, such as those produced in URHICs at low center-of-mass energies.

\section{Gauge invariance in medium}\label{sec4}
Within the VDM, the $\rho$ meson couples to a conserved current, requiring it to be 4-dimensionally transverse, $q_{\mu}\Sigma_{\rho}^{ \mu\nu}=0$. The Ward identities (Eq. \ref{Ward}) must be satisfied by the $\rho\pi\pi$ ($\Gamma_{\mu\,ab3}^{(3)}$) and $\rho\rho\pi\pi$ ($\Gamma_{\mu\nu\, ab33}^{(4)}$) vertices to ensure transversality:
\begin{eqnarray}
\label{Ward}
q^{\mu}\Gamma_{\mu\,ab3}^{(3)}&=&g_\rho \epsilon_{3ab}(D^{-1}_{\pi}(k+q)-D^{-1}_{\pi}(k)) \ ,\\
\label{eq43a}
q^{\mu}\Gamma_{\mu\nu\, ab33}^{(4)}&=&ig_{\rho} (\epsilon_{3ca}\Gamma_{\nu\,bc3}^{(3)}(k,-q)-\epsilon_{3bc}\Gamma_{\nu\,ca3}^{(3)}(k+q,-q)) \ .
\end{eqnarray}
The Ward are trivially satisfied in vacuum, but are broken by the introduction of a pion self-energy. Following Refs.~\cite{Atchison:2023,Urban:1998,Song}, we introduce thermal corrections to the $\rho\pi\pi$ and $\rho\rho\pi\pi$ vertices to restore the Ward identities. The in-medium vertices now take the form:
\begin{eqnarray}
\label{WT3}
\Gamma_{\mu \, ab3}^{(3)}&=&g_{\rho}\epsilon_{3ab}(2k+q)_{\mu}+\Gamma_{\mu \, ab3}'^{(3)} \ ,
\\
\label{WT4}
\Gamma_{\mu\nu \, ab33}^{(4)}&=&2ig_{\rho}^{2}(\delta_{ab}-\delta_{3a}\delta_{3b})g_{\mu\nu}+\Gamma_{\mu\nu \, ab33}'^{(4)} \ ,
\end{eqnarray}
where $\Gamma_{\mu \, ab3}'^{(3)}$ and $\Gamma_{\mu\nu \, ab3}'^{(4)}$ are vertex corrections to the $\rho\pi\pi$ and $\rho\rho\pi\pi$ vertices. 

One can see from Eq. \ref{Trans_conductivity} that only the spacial components of $\Sigma_{\rho}^{\mu\nu}$ influence the conductivity. Therefore, we only address the spacial components of the vertex corrections. For the $\pi N$ interaction we adopt the corrections from  Ref.~\cite{Urban:1998}. Due to the non-relativistic approximations employed in Ref.~\cite{Urban:1998}, the vertex corrections can be written in terms of $\Sigma_{\pi}^{N}$. This greatly reduces the numerical burden when calculating $\Sigma_{\rho}$.  In order to reduce the numerical effort required in our calculation, we propose "effective" vertex corrections for the $\pi\pi$ and $\pi K$ interactions, which are inspired by Ref.~\cite{Urban:1998}. Although these corrections are not directly derived from Lagrangians, they do exactly satisfy the Ward identities. Furthermore, Ref.~\cite{Atchison:2023} found the vertex corrections for the $\pi\pi$ interaction to only provide an $\sim10\%$ increase in the conductivity. Therefore, we do not expect the introduction of simplified vertex corrections to significantly impact the transport peak.

We start by assuming the corrections can be written as a sum of terms proportional to $\Sigma_{\pi}$. For the spacial components of the 3-point corrections, we additionally assume each term is proportional to a momentum $k_{i}$ or $q_{i}+k_{i}$, and can be written as a function of $k$ or $q+k$, as was the case in Ref.~\cite{Urban:1998}. For the 3-point vertex we arrive at the corrections:
\begin{eqnarray}
\label{Eff_VC3_0}
\Gamma_{0 \, ab}'^{(3 \, R\pi)} &=& \frac{g_{\rho} \epsilon_{3ab}}{q_{0}}\Big[\big(\frac{\vec{q}^{2}+\vec{q}\cdot\vec{k}}{(q_{0}+k_{0})|\vec{q}+\vec{k}|}-1\big)\Sigma_{\pi}^{R}(q+k)\nonumber\\
&&+\big(\frac{\vec{q}\cdot\vec{k}}{k_{0}|\vec{k}|}+1\big)\Sigma_{\pi}^{R}(k)\Big]\\
\label{Eff_VC3_i}
\Gamma_{i \, ab}'^{(3 \, R\pi)} &=& g_{\rho} \epsilon_{3ab}\Big[\frac{q_{i}+k_{i}}{(q_{0}+k_{0})|\vec{q}+\vec{k}|}\Sigma_{\pi}^{R}(q+k)+\frac{k_{i}}{k_{0}|\vec{k}|}\Sigma_{\pi}^{R}(k)\Big],
\end{eqnarray}
where, $R= \rho,\sigma,K$, and we have suppressed the 3 in the isospin indices for simplicity. Here, we have divided each term in Eq.~\ref{Eff_VC3_i} by either $k_{0}|\vec{k}|$ or $(q_{0}+k_{0})|\vec{q}+\vec{k}|$, so that the corrections will have units of energy, and each term will be a function of either $k$ or $q+k$. This choice assures the corrections are finite in the limit where $k_{0}$ or $|\vec{k}|$ approach zero, without introducing additional parameters. These restrictions combined with Eq.~\ref{WT3} determine effective 3-point corrections. For the 4-point corrections we assume $\Gamma_{ij \, ab}'^{(4 \, R\pi)}=0$, because its contribution to the transport peak was found to be small in Ref.~\cite{Atchison:2023}. This assumption significantly simplifies the 4-point corrections, and allows us to use Eq.~\ref{WT4} to fix $\Gamma_{i0 \, ab}'^{(4 \, \pi)}$ and $\Gamma_{00 \, ab}'^{(4 \, \pi)}$, giving:
\begin{eqnarray}
\label{Eff_VC4_i0}
\Gamma_{i0 \, ab}'^{(4 \, R\pi)} &=& \Gamma_{0i \, ab}'^{(4 \, R\pi)}=\frac{-ig_{\rho}}{q_{0}}(\delta_{ab}-\delta_{3a}\delta_{3b})\Big[\frac{(-q_{i}+k_{i})}{(-q_{0}+k_{0})|-\vec{q}+\vec{k}|}\Sigma_{\pi}^{R}(-q+k)\nonumber\\
&&-\frac{q_{i}+k_{i}}{(q_{0}+k_{0})|\vec{q}+\vec{k}|}\Sigma_{\pi}^{R}(q+k)\Big]\\
\label{Eff_VC4_00}
\Gamma_{00 \, ab}'^{(4 \, R\pi)} &=& \frac{-ig_{\rho}^{2}}{q_{0}^{2}}(\delta_{ab}-\delta_{3a}\delta_{3b})\Big[2\Sigma_{\pi}(k)-\Sigma_{\pi}^{R}(-q+k)-\Sigma_{\pi}^{R}(q+k)\Big].
\end{eqnarray}
Finally, in Ref.~\cite{Atchison:2023} some vertex corrections were incorporated by resummation into $D_{\pi}$, as an additional pion self-energy $\Sigma_{\pi}^{U}$. These corrections produced approximately 70\% of the increase in the transport peak. It is straight forward include these corrections by adding them to $\Sigma_{\pi}$, when calculating the Landau cut of $\Sigma_{\rho}$. While the addition of $\Sigma_{U\pi}$ violates the Ward identities, this is easily remedied by including additional vertex corrections for $R=U$, as described in eqs. \ref{Eff_VC3_0} through \ref{Eff_VC4_00}.

At the transport peak, the effective vertex corrections are only a few percent smaller than the corrections calculated in Ref.~\cite{Atchison:2023}. However, the two approaches disagree at the $\rho$ pole, where the corrections from Ref.~\cite{Atchison:2023} provide significantly more broadening. Although a more advance approach is necessary for higher invariant mass, the effective corrections appear to reliably approximate the vertex corrections at the transport peak, without the need for cumbersome numerical calculation.  

\section{EM spectral function at finite temperature and density}\label{sec5}
\begin{figure}
\includegraphics[width=16pc]{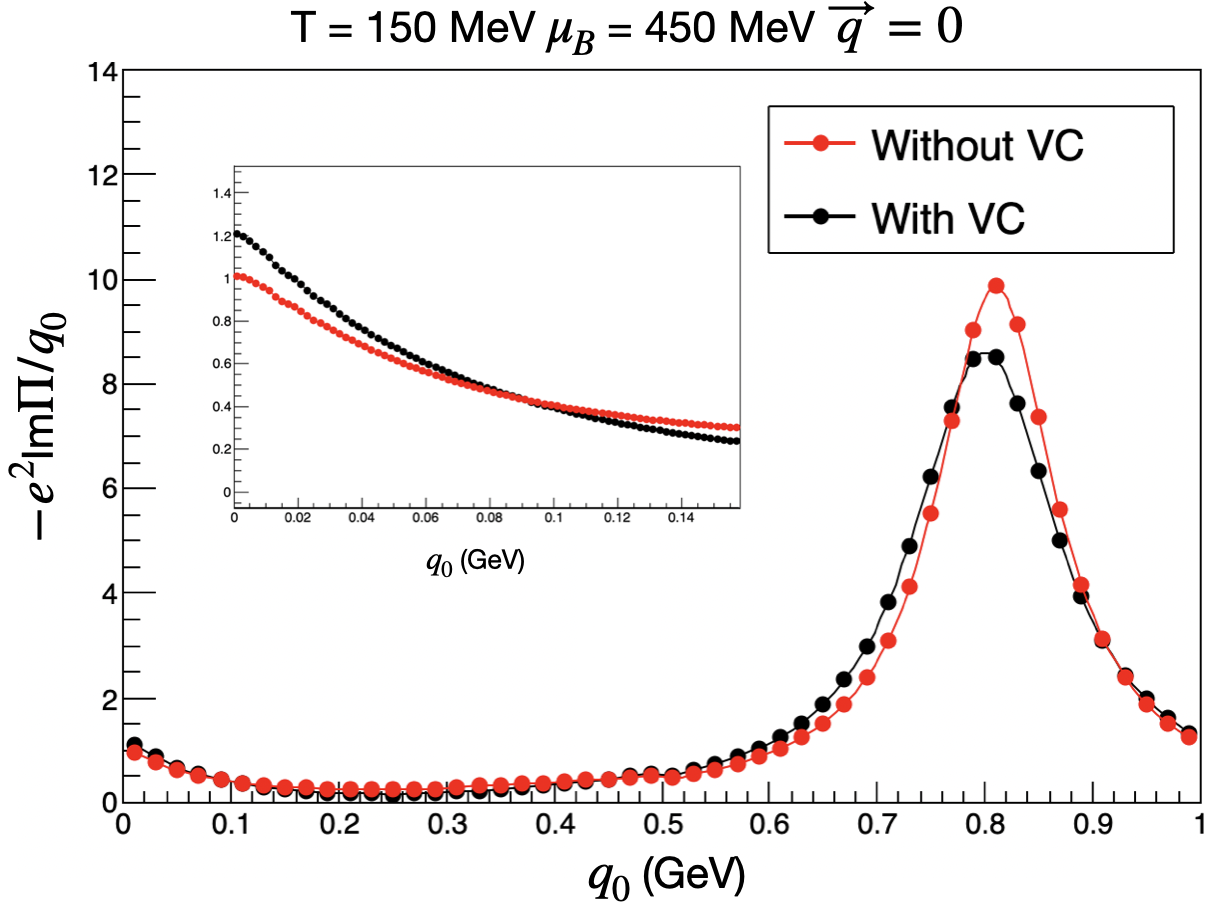}
\includegraphics[width=16pc]{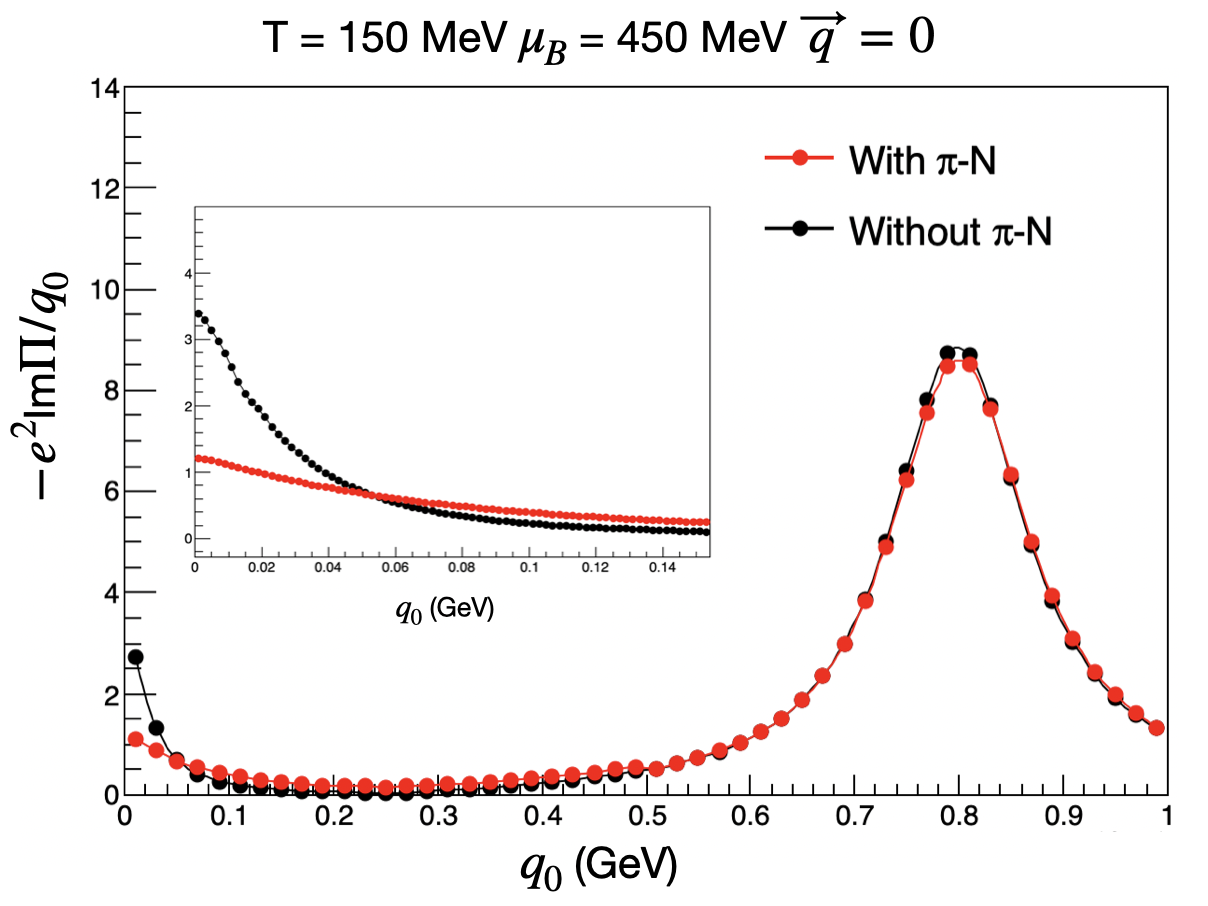}
\caption{The transverse projection of the electromagnetic spectral function at $T=150\textrm{MeV}$ and $\mu_{B}=450\textrm{MeV}$. Left: The spectral function with (black) and without (red) vertex corrections. Right: The spectral function with (red) and without (black) $\pi N$ scatting. The insert emphasises the low energy transport peak. We see that the $\pi N$ interaction causes significant broadening of the transport and mass peaks.} 
\label{Unrealistic_Comp}
\end{figure}
To begin, we examine the effects of the vertex corrections at finite temperature and density. In the left panel of figure \ref{Unrealistic_Comp} we plot the spectral function with and without vertex corrections at $T=150\textrm{MeV}$ and $\mu_{B}=450\textrm{MeV}$. At the $\rho$'s mass peak we observe somewhat less broadening than would be expected from Ref.~\cite{Atchison:2023}, due to our implementation of effective vertex corrections. However, the low energy region agrees well with Ref.~\cite{Atchison:2023}. Here, the vertex corrections increase the height of the transport peak by $\sim15\%$. Our corrections produce a larger enhancement than what was found in Ref.~\cite{Atchison:2023}. This is due to our inclusion of additional vertex corrections for the $\pi N$ and $\pi K$ interactions. Although, we observe an enhancement at the transport peak, the vertex corrections still represent only a modest enhancement.

In the right panel of figure \ref{Unrealistic_Comp} we plot the spectral function with and without $\pi N$ scattering. The $\pi N$ interaction broadens the transport peak, but has little impact on the $\rho$'s mass peak. The lack of broadening of the mass peak is expected because we only dress the $\rho$'s pion cloud, while excluding direct interactions of the $\rho$ with baryons and heavier mesons $(a_{1},\omega, h_{1},K_{1},\pi', \textrm{and}\! f_{1})$. At the transport peak, the $\pi N$ interaction creates substantial broadening, reducing the height of the peak by more than a factor of two. This is because the transport peak is inversely proportional to the pion's width. The introduction of a large nucleon density greatly increases the pion's medium interactions, as seen in figure \ref{PionSelfTotal}. The $\pi N$ scatterings impede the flow of charged pions, broadening the transport peak. This effect could proved beneficially in future experiments aimed at probing the transport peak through low-mass dielectron measurements, as the broadening will push the peak to higher invariant mass~\cite{HADES,STAR:iTPC,ALICE:2022wwr}.

\begin{figure}
\includegraphics[width=16pc]{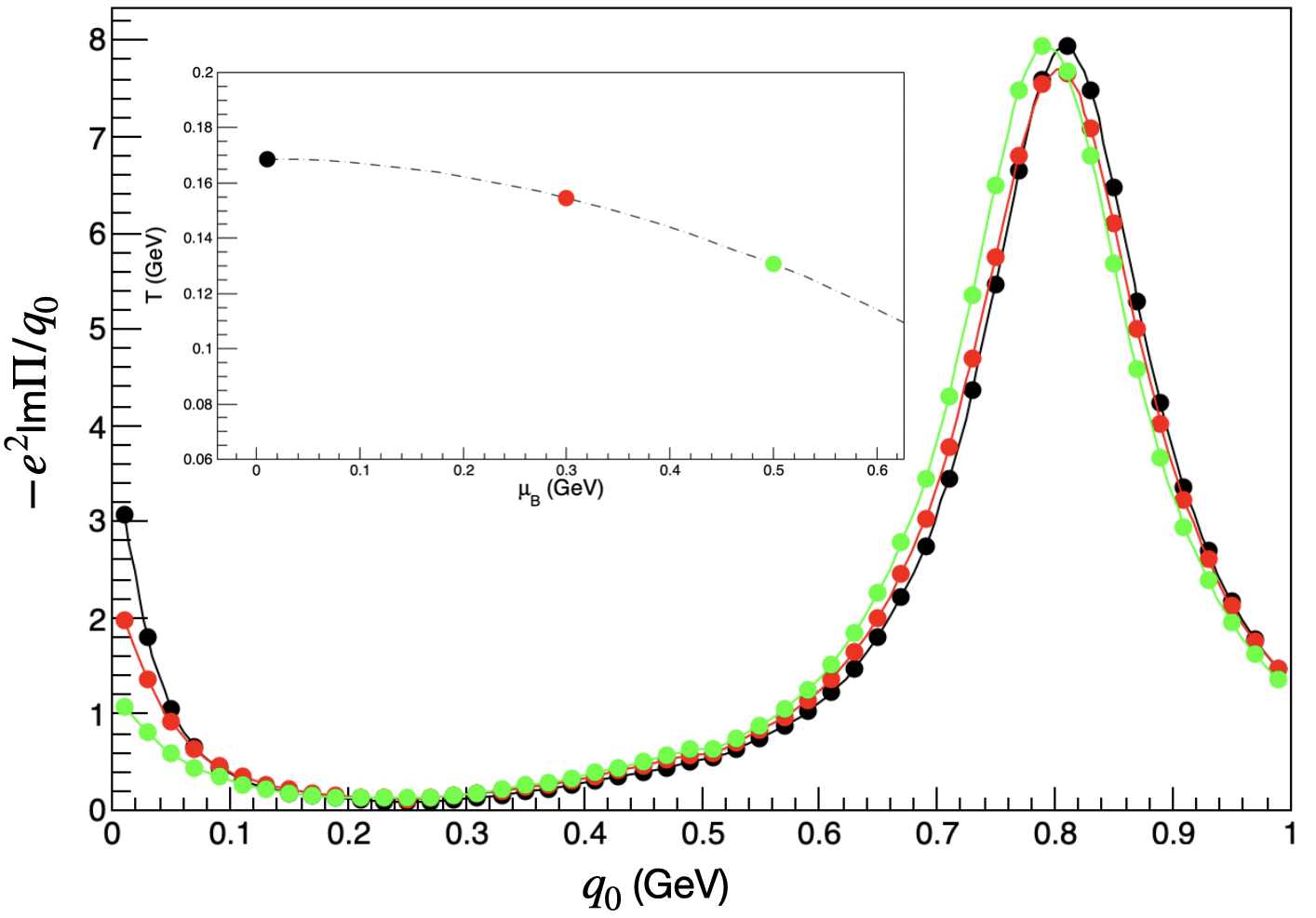}
\includegraphics[width=16pc]{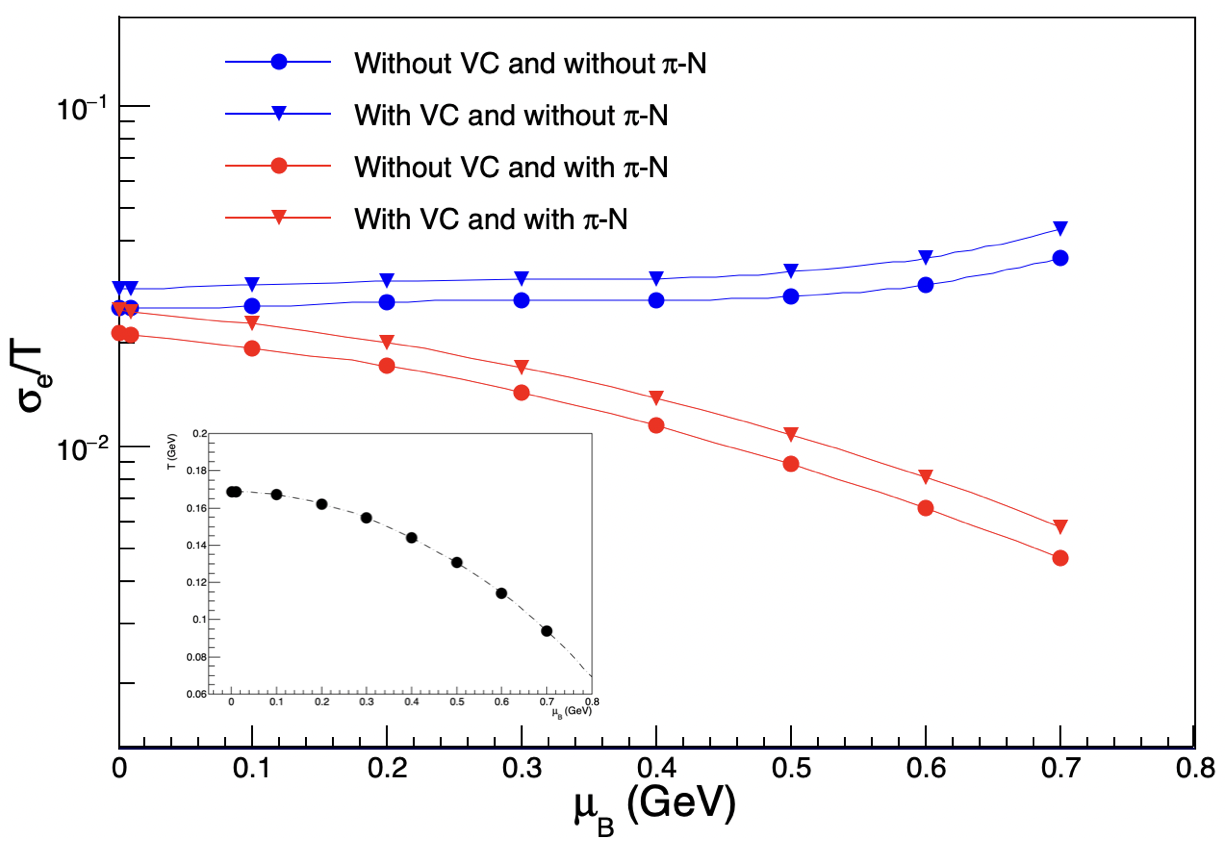}
\caption{Left: The transverse projection of the electromagnetic spectral function for three points along the phase transition line: $\{T=157\textrm{MeV},\mu_{B}=1 \textrm{MeV}\}$ (black), $\{T=145\textrm{MeV},\mu_{B}=300 \textrm{MeV}\}$ (red), and $\{T=115\textrm{MeV},\mu_{B}=500 \textrm{MeV}\}$ (green). Right: The electric conductivity along phase transition line. Red points include $\pi N$ scattering, while blue points do not. Results including vertex corrections are plotted with circles, and results excluding vertex corrections are plotted with triangles. The insert shows the position of each the temperature and baryon chemical potential on the QCD phase transition line.} 
\label{Realistic_Comp}
\end{figure}
Next, we analyse the electromagnetic spectral function peak as a function of temperature and baryon chemical potential. Figure \ref{Realistic_Comp} plots the spectral function at three values along the phase transition line. The phase transition line is estimated from chemical freeze out data at various beam center-of-mass energies via published experimental results~\cite{Becattini_2004,Das_2015}. Following the transition line, we see that as $\mu_{B}$ increases the temperature decreases. The reduction in temperature decreases the pion density and, in turn, $\Sigma_{\rho}$. The transport peak and the $\rho$'s width are produced by the Landau and Unitary cuts of $\Sigma_{\rho}$, respectively. Therefore, in figure \ref{Realistic_Comp} we observe the height of the transport peak drop, and the $\rho$'s mass peak narrow, as we move to lower temperature. 

In the low energy region one might expect the decrease in temperature to weaken the pion's medium interactions, thus narrowing the transport peak. However, while the $\pi$ and $K$ densities decrease with temperature, the nucleon density increases with $\mu_{B}$. As $\mu_{B}$ increases, the contribution from the nucleons more than compensates for the falling $\pi$ and $K$ densities. Therefore, the half-width of the transport peak increases along the transition line. This demonstrates the importance of nucleons at low temperatures and high $\mu_{B}$, as the large nucleon density implies that the $\pi N$ scattering will provide the majority of the $\pi$'s width in this regime.

Finally, we extract the conductivity using Eq.~\ref{Trans_conductivity}. In the right panel of figure \ref{Realistic_Comp} we plot the conductivity over temperature along the phase transition line. We provide results with and without vertex corrections, which provide a $\sim 15\%$ increase. The increase in the conductivity is due to the corrections introducing additional channels that the EM current can couple to the medium through~\cite{Atchison:2023}. Although, this result is somewhat larger than the results from Ref.~\cite{Atchison:2023}, it is consistent with our analysis of the transport peak. 

Results are also shown with and without $\pi N$ scattering. When the $\pi N$ interaction is excluded we see the conductivity rise along the phase transition line. This is due to the temperature dropping as one moves from left to right in figure \ref{Realistic_Comp}. The dropping temperature lowers the $\pi$ and $K$ densities, reducing the $\pi$'s width and increasing the conductivity. Alternatively, when the $\pi N$ interaction is included we observe a falling conductivity over temperature, because the raising $\mu_{B}$ increases the nucleon density. In the non-relativistic limit nucleons do not transport charge. Therefore, increasing the nucleon density increases the $\pi$'s width, without introducing more charge carries. The nucleons create a resistive medium that the charge carrying pions travel through. Therefore, the dichotomy between temperature and $\mu_{B}$ drastically changes the medium from a more weakly interacting pion gas, to a strongly interacting nucleon filled medium.

\section{Summary and future work}\label{sec6}
We have calculated the EM spectral function in hot and dense nuclear matter from a quantum many-body approach, rooted in successful descriptions of thermal dilepton emission spectra in HICs, with the EM current coupling to the medium primarily through the $\rho$ meson~\cite{Rapp:1999us,Salabura:2020tou,Tripolt:2022hhw}. Medium interactions are then introduced through the $\rho$ meson's self-energy. The Landau cut of the $\rho$ meson's self-energy generates the spectral function's transport peak, and conductivity. The Landau cut corresponds to the absorption of a virtual $\rho$ meson by a thermal $\pi$. The $\pi$ and transport peak's widths are subsequently generated by the $\pi$'s interactions with in-medium $\pi$, nucleons, and $K$. We do not consider the direct interactions of the $\rho$ with other mesons or nucleons, because these interactions are heavily suppressed in the low energy regime. In fact, the nucleons do not contribute to first order in the nonrelativistic expansion. Therefore, we find that the $\pi$ acts as the primarily charge carrier in the nuclear medium, while the nucleons and $K$ increase the medium's resistivity.

For the $\pi$'s medium interactions we include $\pi\pi$ S-wave and P-wave scattering through $\sigma$ and $\rho$ resonances, P-wave $\pi N$ scattering through N-hole and $\Delta$-hole excitations, and P-wave $\pi K$ scattering through the $K^{*}$ resonance. We correct any violations of gauge invariance by constructing vertex corrections. For the $\pi N$ interaction we adopt the formalism of Ref.~\cite{Urban:1998,Urban:1999im}. For the $\pi\pi$ and $\pi K$ interactions we construct effective vertex corrections that, although not derived from Feynman diagrams, exactly maintain the Ward identities. While the vertex corrections provide a modest $(\sim15\%)$ increase to the transport peak and conductivity, the $\pi$'s width is the most important factor in generating the transport peak. For example, the addition of a sizable nucleon density at finite $\mu_{B}$ significantly increases the $\pi$'s width and lowers the conductivity. This feature is critical when examining the EM spectral function and conductivity along the phase transition. The rising $\mu_{B}$ results in a broadening transport peak and falling conductivity, in stark comparison to the more QED like rising conductivity found for the pion gas. Our results suggest that in the low temperature high $\mu_{B}$ limit of the QCD phase diagram, pions at as the primary charge carrier, while nucleons serve as a approximately stationary medium providing resistance.

Experiments underway at GSI, RHIC, and the LHC plan to measure very-low-mass dileptons, to as low as tens of MeV ~\cite{HADES,STAR:iTPC,ALICE:2022wwr}. The large broadening of the transport peak at finite $\mu_{B}$ suggests that these efforts have the potential to measure the electrical conductivity along phase transition line by accessing the low-energy dilepton spectrum. To this ends, we intend to carry our model to finite momentum, in order to calculate thermal dilepton emission rates at low-invariant mass. This will require analysis of the effective vertex corrections, to ensure they are reliable at finite momentum. Finally, although the nucleons do not contribute to the conductivity at first order in the nonrelativistic expansion, for low temperature and high $\mu_{B}$ next to leading order effects may not be negligible. Therefore, the size of these effects under experimental conditions should be investigated.

\section*{Acknowledgements}
J.A.~was partially supported by the U.S.~National Science foundation under grants PHY-1913286 and PHY-2209335.

F.G.~is supported in part by the U.S.~Department of Energy Office of Science under grant No.~DE-SC0005131.

\bibliographystyle{elsarticle-num} 
\bibliography{mybib}





\end{document}